\documentclass[aps,prl,amsmath,twocolumn,amssymb,titlepage]{revtex4-1}

\usepackage[utf8]{inputenc}
\usepackage[dvipsnames]{xcolor}
\usepackage{graphicx}
\usepackage[colorinlistoftodos]{todonotes}
\usepackage{soul}
\usepackage{url}

\newcommand{\sklearn}{\textsc{scikit-learn} }

\newcommand{\SHAP}{\textsc{SHAP} }

\newcommand{\imagenet}{\textsc{imagenet} }
\newcommand{\alphafold}{\textsc{alphafold}}

\begin{document}

\title{Interpretable and Explainable Machine Learning for Materials Science and Chemistry}

\author{Felipe Oviedo \textdaggerdbl}
\email{foviedo@alum.mit.edu}
\affiliation{Massachusetts Institute of Technology, Cambridge, MA 02139, USA}
\affiliation{Microsoft AI for Good Research Lab, Redmond, WA 98052}

\author{Juan Lavista Ferres}
\affiliation{Microsoft AI for Good Research Lab, Redmond, WA 98052}

\author{Tonio Buonassisi}
\affiliation{Massachusetts Institute of Technology, Cambridge, MA 02139, USA}

\author{Keith T. Butler \textdaggerdbl}
 \email{keith.butler@stfc.ac.uk}
\affiliation{%
 SciML, Scientific Computing Department, Rutherford Appleton Laboratory, Didcot OX110D, UK\\
}
\affiliation{%
Department of Chemistry,
University of Reading,
Reading, RG6 6AD, UK
}

\author{\textdaggerdbl Equal contribution.}

\date{October 2021}

\maketitle

\section{Conspectus}

Machine learning has become a common and powerful tool in materials research. As more data becomes available, with the use of high-performance computing and high-throughput experimentation, machine learning has proven potential to accelerate scientific research and technology development. While the uptake of data-driven approaches for materials science is at an exciting, early stage, to realise the true potential of machine learning models for successful scientific discovery, they must have qualities beyond purely predictive power. The predictions and inner workings of models should provide a certain degree of explainability by human experts, permitting the identification of potential model issues or limitations, building trust on model predictions and unveiling unexpected correlations that may lead to scientific insights. 
In this work, we summarize applications of interpretability and explainability techniques for materials science and chemistry and discuss how these techniques can improve the outcome of scientific studies. We start by defining the fundamental concepts of interpretability and explainability in machine learning, and making them less abstract by providing examples in the field. We show how interpretability in scientific machine learning has additional constraints compared to general applications. Building upon formal definitions of interpretability in machine learning, we formulate the basic trade-offs between the explainability, completeness and scientific validity of model explanations in scientific problems. In the context of these trade-offs, we discuss how interpretable models can be constructed, what insights they provide, and what drawbacks they have. We present numerous examples of the application of interpretable machine learning in a variety of experimental and simulation studies, encompassing first-principles calculations, physicochemical characterization, materials development and integration into complex systems. We discuss the varied impacts and uses of interpretabiltiy in these cases according to the nature and constraints of the scientific study of interest. We discuss various challenges for interpretable machine learning in materials science and, more broadly, in scientific settings. In particular, we emphasize the risks of inferring causation or reaching generalization by purely interpreting machine learning models and the need of uncertainty estimates for model explanations. Finally, we showcase a number of exciting developments in other fields that could benefit interpretability in material science problems. Adding interpretability to a machine learning model often requires no more technical know-how than building the model itself. By providing concrete examples of studies (many with associated open source code and data), we hope that this account will encourage all practitioners of machine learning in materials science to look deeper into their models.

\section{Introduction}
\textit{"Where is the knowledge we have lost in information?"}
\newline
\newline
The lamentation on the modern condition in the opening stanza of T.S. Eliot's \textit{The Rock} could just as appropriately, if more prosaically, be used to summarise much of the scepticism of scientists towards machine learning (ML) as applied to traditional scientific subjects. In a plenary lecture at a recent international conference, one leading researcher in theoretical chemistry remarked "[a]t least 50\% of the machine learning papers I see regarding electronic structure theory are junk, and do not meet the minimal standards of scientific publication", specifically referring to the lack of insight in many publications applying ML in that field. But is scientific knowledge inevitably lost in machine learning studies, if not how can it be extracted and how does this apply to machine learning in the context of scientific research? 
In this perspective, we set out to provide some answers to these questions.

There have already been numerous efforts to build a taxonomy of interpretability and explainability methods for machine learning models, two noteworthy examples that we draw upon for explaining classical ML models and deep neural networks are references \cite{molnar2019} and \cite{gilpin2018explaining} respectively. Both of these references provide an excellent in-depth comprehensive review of different methods. In the \textit{Key Concepts} section, we provide a brief overview of some of the concepts that will be most important for following the rest of this account, but we recommend these references for readers interested in learning more. 

During the account, we draw in the experience of the authors in using machine learning for understanding and guiding experiments, and enhancing theory and simulation to give a broad overview of how interpretability and explainability have a role to play across the materials science disciplines. We cover a range of methods, starting from building inherently interpretable models and then introducing techniques for extracting explanations and interpretations from models. Many of the methods we highlight for interpretability are easily implemented using existing software, meaning that we believe that there is no reason why ML applied to materials science should remain a black box. Moreover, we also consider some frontiers in interpretable ML models, which mean that far from obfuscating, ML models offer the promise of new physical insights. We can retrieve and perhaps even expand the knowledge latent in the vast amounts of information currently available in materials science. In addition to this promise, we discuss the potential roadblocks and particular challenges for machine learning interpretability in the field.

\section{Key concepts}

\begin{figure}[h!]
   %\begin{center}
   \includegraphics[width=\linewidth]{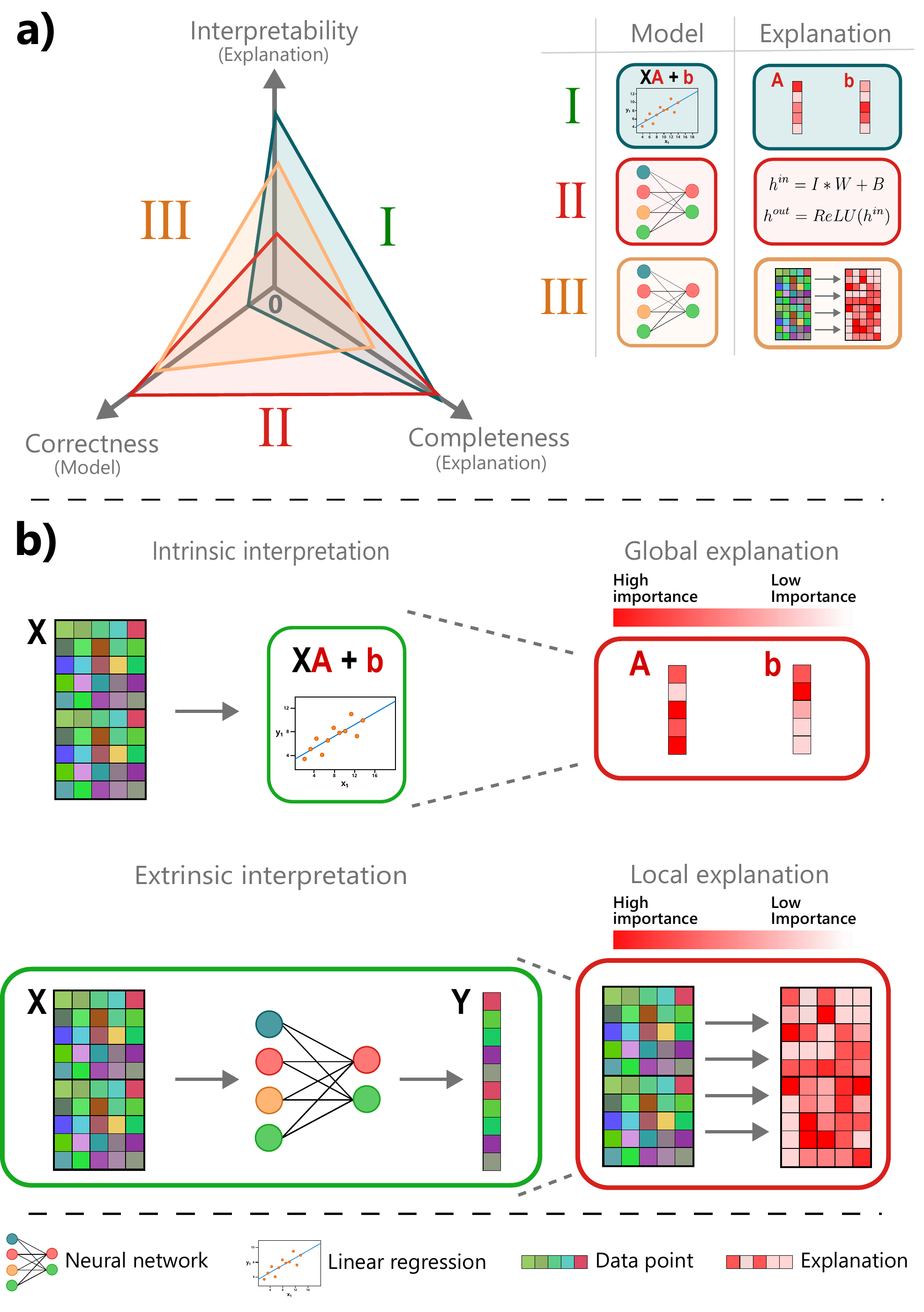}
   %\end{center}
   \caption{\textbf{Key concepts} a) Any explanation of a machine learning model has some inherent trade-offs to it. In particular, any explanation should balance completeness, \textit{i.e.} how well the given explanation approximates the operating mechanisms of the actual model and interpretability, \textit{i.e.} how well can a human subject understand the given explanation. In models that approximate real-world phenomena, we argue that a third dimension exists: correctness \textit{i.e.} how correct a given explanation from the physical or chemical points of view. In the text, we explain the various trade-offs in more detail in examples I, II and III. b) Illustration of local/global explanations and intrinsic/extrinsic interpretability. In I) a linear model is a intrinsic interpretable model. By construction, the vector coefficients A and b can be interpreted directly, as represented by the color scale. In the same way, a linear model can be explained globally as both A and b are applied to all inputs X and have constant contributions for each input. II) A neural network requires extrinsic explanations. Due to its non-linear nature, interpretations of the model require observing the inputs along with the outputs. In the same way, a neural network model is better explained using local explanations: each input interacts in a different way with the model to generate specific outputs.} \label{fig:key-concepts}
\end{figure}

Interpretability of machine learning models is at the forefront of research in computer science; as such there is an abundance of technical jargon associated with the subject. We try to eliminate unnecessarily technical explanations in this account, but in the interests of avoiding the quandary of being ``divided by a common language''\footnote{The original quote about England and America being ``divided by a common language'' is variously attributed to Oscar Wilde, George Bernard Shaw and Winston Churchill.} we start by clarifying some of the terms we see as unavoidable for a proper exploration of the subject.

\textit{Classical/Deep machine learning}. When we refer to classical methods we mean any ML method that is not neural network-based, when we refer to deep learning (DL) methods we are referring to any method based on neural network architectures. This definition is important because there is generally a difference in how we interpret classical or deep learning methods. Classical methods are generally trained on structured data, with human-defined and (more-or-less) interpretable features, for example in materials a feature could correspond to the mean electronegativity of elements in a compound \cite{molnar2019, lipton2018mythos}. Deep learning methods, on the other hand, are trained on less structured data, for example images or text corpora. Deep learning methods learn reduced dimensionality representations of these unstructured inputs (known as representation learning \cite{goodfellow2016deep}) and then use these learned (as opposed to prescribed) features to perform non-linear regression or classification. Because of this difference in how features are developed and used, interpreting classical and deep models often requires different approaches.

\textit{Interpretability/explainability/completeness}. In much of the literature on ``interpretable'' and ``explainable'' ML, the two terms are used almost interchangeably. However, based on the clear differentiation presented in \cite{gilpin2018explaining}, we follow the definition that interpretability is a necessary but not sufficient condition for explainability: the missing ingredient is completeness. A model is interpretable if it provides explanations about its mechanics. Completeness is concerned with how accurately this explanation reflects the actual operation of the model. In principle, a complete description of a  model is always possible: for example, a deep learning model can be \textit{completely} explained by its mathematical operations, but this is of little interpretability to a human user \cite{gilpin2018explaining}. In a chemistry analogy, one may interpret a reaction energy calculated using quantum mechanics in terms of the frontier orbital energies of reactants and products, while this may be a useful interpretation, it is clearly not a full explanation, \textit{i.e.} it is lacking completeness. 

In practice, in model explanations there is often a trade-off between completeness and interpretability and it is important that practitioners be aware of this tension, we represent this tension in a radar plot in Figure~\ref{fig:key-concepts}. In Figure~\ref{fig:key-concepts}, the neural network of example II can be explained by a complete mathematical description of its operations, but it sacrifices interpretability compared to the less complete attribution map of explanation III (these maps are described in section~\ref{sec:deep-interpretations} \textit{Deep Interpretations}).  

\textit{Correctness/Causation}. While the computer science literature on interpretability tends to concentrate on the completeness \textit{vs} interpretability trade-off, we introduce an additional factor for consideration, particularly in scientific applications; correctness. According to the famous aphorism, ``all models are wrong, but some are useful'', correctness is concerned with the degree of scientific wrongness of a given model and explanation. In scientific models, there is often a tension between how faithfully a model reproduces measurements and its complexity, as depicted in the radar plot of Figure~\ref{fig:key-concepts}. In the figure, example I presents a linear model which has a high degree of completeness and interpretability by definition, but is may be limited in terms of adequately capturing physical or chemical phenomena. Thus, explanations of complex neural network models may allow a higher degree of physical correctness, as shown in examples I and II. Conversely, in simulations, an electronic structure calculation may provide highly accurate estimates of the bulk modulus of a solid, but a less accurate pair potential model may provide more intuitively understandable results, this is because the parameters in the electronic structure model are highly abstract representations of electron density, while the pair potential is based on simple heuristics for the different forces between atoms. In this scenario, explanations of each model will provide different degrees of correctness: explanations of the electronic structure model will be inherently more correct, but may lack interpretability or completeness depending on the explanation technique employed. Additionally, more complex and correct models often become intractable for all but the simplest systems, enforcing a resort to simpler representations and explanations. In all cases the correct choice depends on the motivation for building the model, and how important interpretability and completeness are compared to physical or chemical correctness. 

An important consideration is that, although a 'correct' ML model may approximate a physical phenomenon and also give a rough idea of cause and effect, \textbf{this does not translate necessarily into the discovery of \textit{'causation'}}, unless there is specific experimental setup (controlling for confounding and noise) or a particular assumptions regarding the phenomenon. Confusing interpretability and causality is a common issue in scientific machine learning, which is in part caused by the ambiguity of the concept of an 'explanation': explaining a model of a physical phenomenon will give an idea of the predictive power of variables, but is not the same as giving a causal explanation of the physical phenomenon. Hence, we argue that, in order to provide physical insights, machine learning explainability techniques have to be supplemented by adequate follow-up experimentation or explicit causal modelling. In this perspective, we consider interpretable machine learning as a useful tool to generate scientific hypotheses and understand model predictions, however these hypotheses need to be later confirmed by the mentioned techniques.

\textit{Local/global explanations.} Local explanations tell us why a model reached a certain decision for a given case or data point, global explanations tell us why the model generally behaves as it does. In Figure~\ref{fig:key-concepts} the coefficients of the linear regression provide a global explanation of model behaviour, while the salience map provides a local explanation. From classical thermodynamics, the equation
\begin{equation*}
    \Delta G = \Delta H - T \Delta S
\end{equation*}
provides a global explanation of the relationship between free energy, enthalpy, temperature and entropy, while an individual calorimetry experiment can give a local explanation on how the entropy of a particular system depends on temperature. We note that in some cases, it is also possible to aggregate the results of local explanations to generate a global explanation \cite{lundberg2020local}.

\textit{Intrinsic/extrinsic interpretations.} Interpretability can come from examining the model itself, or alternatively by examining how the model responds to stimuli; the former is an intrinsic interpretation, the latter is an extrinsic interpretation. Intrinsic methods generally provide global explanations, because the interpretation depends on the construction of the model, intrinsic methods are specific to given types of ML algorithms. The canonical example of an intrinsically interpretable model is linear regression, as illustrated in Figure~\ref{fig:key-concepts}. Intrinsic interpretations for classical ML methods are very well-established and have been developed over several decades \cite{molnar2019}. Extrinsic methods, on the other hand, are generally model-agnostic or exploit specific inductive biases (prior assumptions) in models and often provide local rather than global explanations, because they rely on perturbations of the input data and observation of how the model responds. Because deep learning methods rely on huge numbers of learnable parameters (routinely in the millions) and non-linear transformations, it is unlikely that one could intrinsically examine the model as it is and understand how it works. Therefore, interpretability of deep learning methods comes from extrinsic methods. Many extrinsic methods developed for classical ML are also applicable to deep learning cases. However, because the descriptors of deep learning models are learned inside the model, rather than provided, special methods for uncovering these features are needed. For this reason a range of deep-learning-specific interpretation methods have also been developed \cite{molnar2019, lipton2018mythos}.

\section{Intrinsically interpretable models\label{intrinsic-examples}}

A range of atomistic ML models have been introduced in recent years. The focus has mainly been on the regression of atom-resolved properties, or global properties as dependent on individual atomic environments. The construction of structural descriptors is often guided by physical ideas, encoding information about environments and symmetries, but this is not an indispensable practice, as complex neural networks have also been used to capture materials structures from raw data inputs. The former naturally lend themselves to interpretable models and indeed have been used to reduce the complexity of structure-composition spaces and draw interesting conclusions from large datasets \cite{nicholas2020understanding}. The development of physically motivated interatomic potentials from machine learning has been comprehensively covered in other review articles\cite{deringer2019machine}.

An alternative approach to building atomistic models is to tabulate a wide range of physical descriptors of a material's composition and structure and to use machine learning to discover relationships between descriptors and properties. Facilitated by data sets such as Materials Project/AFLOW/OQMD/ \cite{saal2013materials, calderon2015aflow, Jain2013} etc., it is relatively straightforward for a researcher with a basic knowledge of Python to obtain a set of materials with associated properties of interest. Packages such as Matminer/Magpie \cite{ward2016general, ward2018matminer}, make it easy to build descriptors for the materials. These descriptors generally consist of combinations of means, sums and standard deviations of elemental properties (\textit{e.g.} electronegativity, number of valence electrons and so on). After appropriate pre-treatment the resulting vector of properties is the input for a classical ML model such as linear regression, decision tree, or more sophisticated ensemble versions of these for example XGBoost, and the model is fitted to reproduce the target property.

It might seem that this is not necessarily an intrinsically interpretable approach (indeed methods related to these models are also discussed in the section on extrinsic interpretations), however in many classical ML methods it is possible to examine the model to see how features contribute to the predictions. Linear and generalized linear models provide direct interpretations by analyzing fitted coefficients, along with their confidence bounds. Tree-based models are interpretable because the order and threshold of decisions executed by the tree to reach an answer can be observed, even when trees are used in ensembles (such as random forests or boosted trees) the degree to which a given parameter splits the data can be obtained and is linked to how important that parameter is for the final prediction; thus providing an interpretation of feature importance. This kind of approach has been been used for example to show which features affect the band gap of a material, or the dielectric response \cite{takahashi2020machine, davies2019data}. Support vector machine methods are also interpretable using similar feature importance inspection. 

While feature importance scores can offer insights, they can be misleading. The methods used within many widespread machine learning packages, such as \sklearn \cite{scikit-learn} have some well-known pathologies. These kinds of feature importance metrics tend to favour continuous over categorical features and care should be taken in particular when using categorical features with high dimensionality or continuous features with wide ranges \cite{strobl2007bias}. In materials science the features that we use are often of vastly different ranges and categorical dimensions, which means that the feature importance obtained by default from these decisions trees should be treated with great caution, particularly where counter-intuitive results are obtained. In the same way, the actual importance of features and their trends can be masked by observed confounding by other features.

Sometimes interpretability can be improved by reducing the number of features, while minimally affecting performance. It is possible to use regularisation techniques to limit the number of descriptors to those that are most important for capturing the relationship between the data and the property of interest. Regularisation approaches such as SISSO (sure independence screening and sparsifying operator) and LASSO (least absolute shrinkage and selection operator), as well as approaches based on perturbing features and retraining models have all been used to to produce ostensibly more interpretable models. Regularisation approaches have been used to re-appraise structure prediction heuristics, for perovskites and zinc blende/wurtzite systems \cite{ouyang2018sisso}. LASSO has also been used to identify important factors for predicting dielectric breakdown thresholds in materials \cite{kim2016organized}. In an example of perturbation methods, backward feature elimination was applied to identify four descriptors most important for predicting superconductor critical temperatures \cite{stanev2018machine}. 

While feature regularisation can be useful for constructing lower dimensional models, they are not without their limitations and potential pitfalls, in particular in the presence of correlated features. For example in LASSO-type 
methods, if a group of features are highly correlated LASSO often arbitrarily chooses one feature at the expense of the others in the group. In perturbation elimination methods, high levels of correlation mean that if an important feature is dropped from the model it may be compensated for by a correlated feature, thus masking the importance. In general, it is good practice to use feature elimination approaches in conjunction with correlation metrics, for example Pearson or Spearman correlation metrics, although one should also remain vigilant as low correlation scores do not necessarily mean unrelated features.

We finish dealing with intrinsically interpretable models by noting that it is also important not to fetishize simpler models in the name of interpretability. Particularly important in this regard is the scenario of model mismatch, where the model form fails to capture the true form of a relationship \cite{lundberg2020local}, \textit{i.e.}, according to our previous definition, provides low correctness. For example, if a linear model is used to capture a non-linear relationship, the model will increasingly attribute importance to irrelevant features in an attempt to minimise the difference between the model predictions and the training data and will ultimately produce meaningless explanations. In machine learning literature, a common solution to preserve predictive power and allow high intrinsic interpretability is using generalized linear models with specific linkage functions or generalized additive models (GAMs) \cite{nori2019interpretml}. For example, GAMs have been used to model and interpret the driving factors of chemical adsorption of subsurface alloys \cite{esterhuizen2020theory}, modelling a non-linear process with a high-degree of interpretability.

\begin{figure}[h!]
   %\begin{center}
   \includegraphics[width=0.7\linewidth]{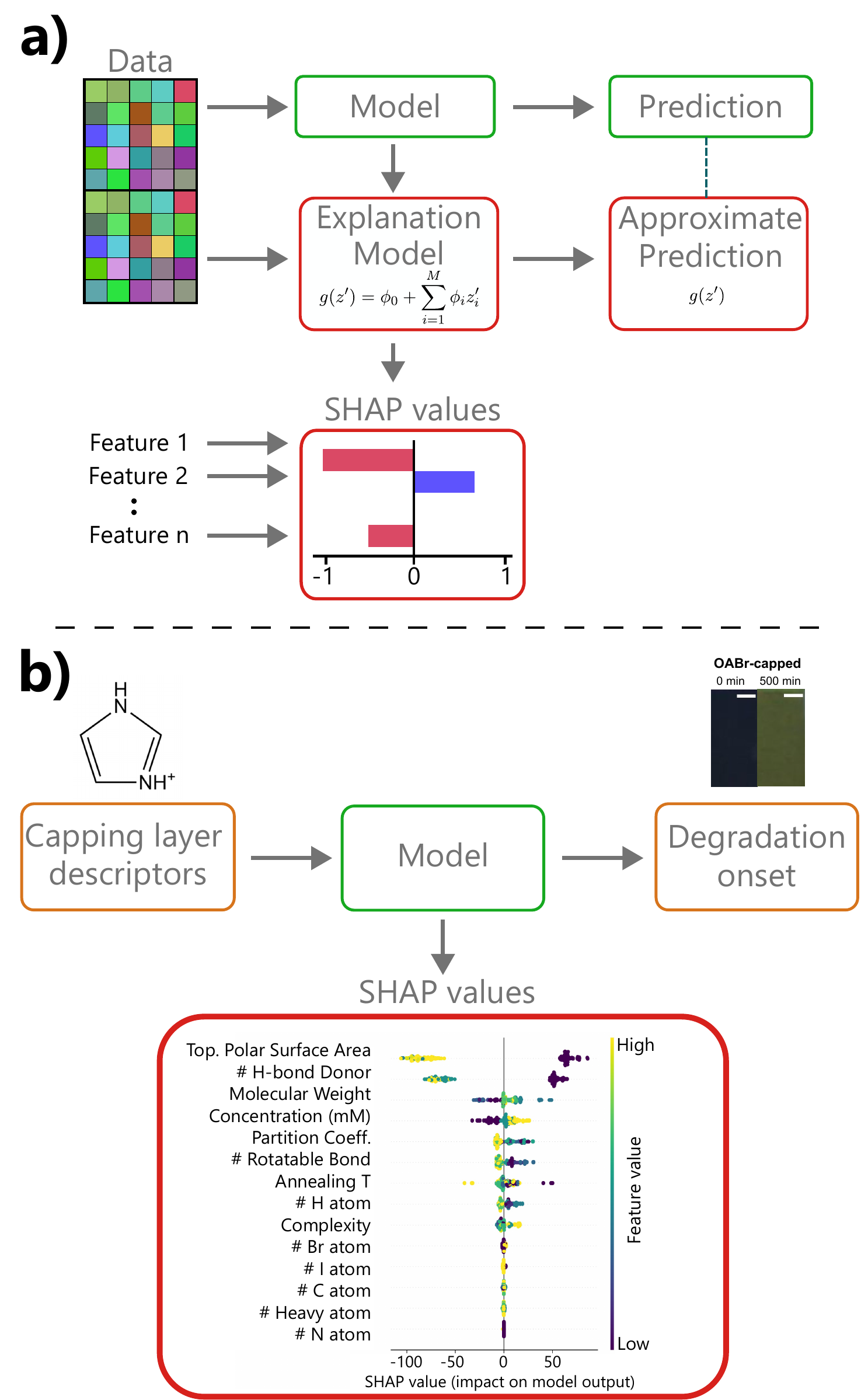}
   %\end{center}
   \caption{\textbf{Interpretability with \SHAP values} (a) \SHAP values are a generalization of various black-box explainability methods. \SHAP values work by approximating the output of a model with a local linear explanation model. The coefficients of this explanation model quantify the local effect of each feature on the output. The coefficients can be aggregated to get global feature contributions. (b) Case study of \SHAP analysis in material science \cite{hartono2020machine}. A model is built to relate the physio-chemical descriptors of a capping layer of lead halide perovskite solar cells. Then, a machine learning model is trained to predict the onset time of degradation of the solar cells under ambient conditions. \SHAP analysis allows to identify the dominant descriptors in the model, shown in the figures as a distribution of local \SHAP values. \textit{Top polar surface area} and \textit{H-bond donor} have the most significant impact the output's prediction and are demonstrated to have dominant importance by additional experimentation. \label{fig:shap}}
\end{figure}

\section{Model Interpretation Methods}

While some ML methods offer intrinsically interpretable results, many more complex models such as deep neural networks (DNNs) are not as easily understood. Also, even when models are inherently interpretable by examining feature importance, extrinsic interpretation methods can provide additional insights impossible by examining the model alone. We begin this section by considering model-agnostic methods for an explanation, we later consider methods that specifically work for  DNNs. Within both model-agnostic and DNN-specific we can have local or global explanations.

\subsection{What-if interpretations \label{what_if}}

 There are a range of ``what-if'' analysis approaches that work by examining how the value of the model output changes when one or more of the input values are modified. Partial dependence plots (PDPs) examine how changing a given feature affects the output, ignoring the effects of all other features\cite{friedman2001greedy}, for example we could look at the effect of the mean atomic mass of a material on the dielectric response, marginalising all other factors using a model such as that presented in reference \cite{takahashi2020machine}. One drawback of marginalising all other features is that confounding relationships are missed and can mask effects, for example imagine increased mean atomic mass increased the dielectric response in a dense material, but decreased it in a porous material, these factors would cancel in PDP.  Individual conditional expectations (ICE) plots are closely related to PDPs, but overcome this limitation and allow group factors to be uncovered \cite{goldstein2015peeking}. PDP and ICE analysis have been widely applied in fields where the application of ML and informatics are significantly more mature than in materials science for example genetics, but they are surprisingly under-utilised in materials science. 

Applying \SHAP analysis on support vector regression (SVR) model it was possible to understand how physical descriptors contribute to the model's ability to predict the dielectric constant of crystals revealing relationships similar to long established empirical models, but with greater predictive power \cite{morita2020modelling}. \SHAP values are also now being more commonly applied to give global as well as local explanations - for example in models that predict atomic charges \cite{korolev2020transferable}. \SHAP analysis can also be applied to neural network models, where input vectors are handcrafted descriptors, this kind of analysis has recently been used to extract chemical rules for polymer composition property relationships and to identify important factors for controlling nano-particle synthesis \cite{kunneth2020polymer, mekki2020two}.

\SHAP analysis is increasingly being embraced by the materials science community, and we believe that this type of interpretation could and possibly should become routine for models where handcrafted physical features are used. However there are limitations to \SHAP analysis related to causation and correlation, which should be considered when applying it. First, like all of the what-if analyses presented Shapley values can sample unrealistic combinations where parameters are correlated, for example in a material an input combination where the HOMO energy is higher than the LUMO energy could be explored despite being physically unrealistic. Second Shapley values are arrived at by including parameters in sequence, but there is no notion of how one feature may directly cause another, so having a low HOMO may be causally related to having a large band gap, but the Shapley value will be calculated as if neither of these features is causally related to the other. We consider some possible solutions to these limitations in the "Physical Knowledge Beyond Model Explanation" session.

\subsection{Deep interpretations}\label{sec:deep-interpretations}

\begin{figure*}
   %\begin{center}
   \includegraphics[width=0.9\linewidth]{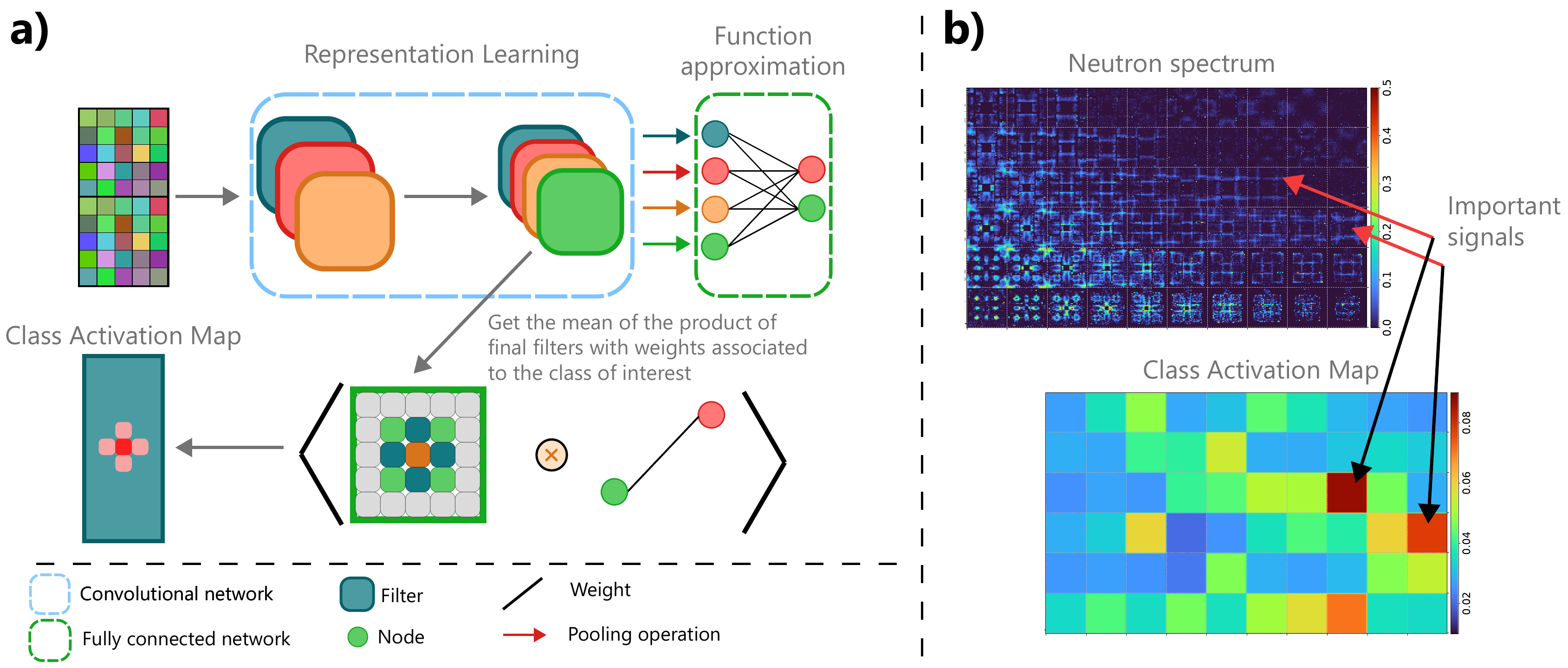}
   %\end{center}
   \caption{ Interpreting the results from a deep convolutional neural network. a) Schematic of class activation maps (CAM): The trained model is presented with a new instance which is passed through the network. The filters in the final convolutional layer are pooled to a single node, the weights connecting this node are multiplied by the filter and the average of all resulting weighted filters is taken and projected back onto the original image, to show the important regions for the classification. b) CAM in action - a CNN was trained to classify magnetic Hamiltonians based on inelastic neutron scattering spectra (upper) and highlight the regions of energy transfer in Q-space that are important for making distinctions using a CAM (lower). The regions identified by the CNN/CAM match with the regions that a trained physicist identifies, but in a fraction of the time\cite{butler2020interpretable}.\label{fig:extrinsic-nn}}
\end{figure*}

As we described in the ``Key concepts'' section, deep learning methods learn non-linear representations rather than relying on handcrafted inputs, because of this there are a number of DL-specific methods for interpretability. Interpretation methods for DL models typically take one of two approaches, \textit{processing} methods or \textit{representation} methods \cite{gilpin2018explaining}. Processing methods examine how the model processes a given input in similar way to a what-if analyses, while representation methods attempt to interpret learned representations or intrinsically learn representations that have some degree of interpretability.

The most popular approach to understanding how the DL model processes data are salience methods. Salience methods are exemplified by early work where networks are repeatedly tested with the same input image, but with different regions blocked out, to determine which areas contribute most to classification \cite{zeiler2014visualizing}. Since the early efforts in this area, a range of methods have been developed which examine a balance between the areas of the network which which respond most strongly to a given input (the activations) and the areas which are most sensitive, \textit{i.e.} where changes in activations would change the output most (the gradients). An overview of various methods for salience mapping is available elsewhere \cite{ancona2018better}. The class activation map (CAM)/grad-CAM \cite{zhou2016learning, selvaraju2017grad} approach builds a map of the input regions that are responsible for a classification by calculating how the different convolutional filters contribute to that classification and building a weighted average of these activations, which can then be projected onto the input image, the operation of CAM is presented schematically in Figure~\ref{fig:extrinsic-nn}. 

\begin{figure}
   %\begin{center}
   \includegraphics[width=\linewidth]{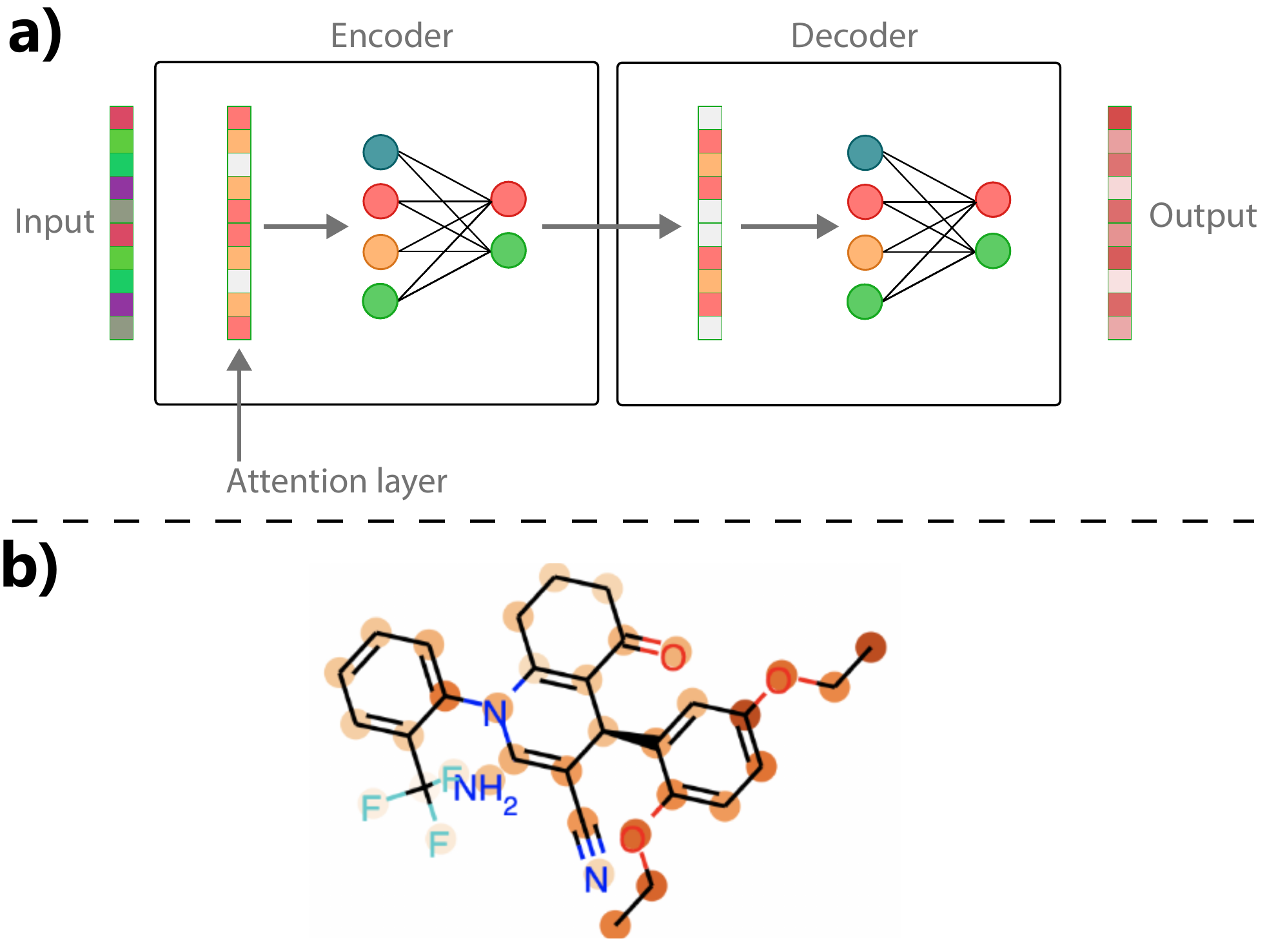}
   %\end{center}
   \caption{ a) The attention mechanism in a transformer neural network. The network learns to translate between sequences of arbitrary length, through encoder and decoder networks. The encoder network takes an embedding of vector of each element of the input and passes it through an attention layer, which learns how much to correlate different members of the input sequence, this then goes through a standard multi-layer perceptron and can be repeated an arbitrary number of times. b) Attention-derived map of the molecular fragments found to be important for predicting hydrophobicity in reference~\cite{payne2020bert}\label{fig:transformer-nn}}
\end{figure}

Representation methods are also very popular approaches to deep learning interpretability.
Recently, transformer architectures have demonstrated outstanding performance on a variety of vision and language tasks. Transformers utilize attention mechanisms\cite{vaswani2017attention}, which learn a weighted masking of different sections of the input data during an encoding procedure see Figure~\ref{fig:transformer-nn}. The transformer begins by learning an ``embedding'' for each element of the input, a simple example could be a vector for an atom of the length of all other elements in the periodic table, the embedding then represents how a given atom is related to all other atoms. This embedding then passes through the attention layers, which learn how much attention elements of the vectors pay to each other. These representations, known as \textit{attention masks}, can be interpreted in similar way to salience maps, and determine sections of the input data that a model exploits for making predictions. Commonly during training, an attention-based model learns multiple attention masks from the training data. These attention masks can be aggregated in various ways to extract regions that are important for achieving the model's tasks \cite{wang2021crabnet, wang2021compositionally}. For example, in \cite{payne2020bert}, a language model trained on tokenized molecular structures was able to correctly identify the areas in the molecule that are active sites for reactions, see Figure~\ref{fig:transformer-nn}. In a similar way, the authors of a transformer model trained on chemical reaction data were able to perform atom-mapping and learn chemical grammars \cite{schwaller2021extraction}, \textit{i.e.} identify atoms during a chemical reaction, by interpreting its learned attention map. Attention maps of composition vectors revealed that in a predicting bandgaps of Si containing materials, Si ``paid attention'' to n-type dopants for predicting the system bandgap, within one of the attention mechanisms\cite{wang2021compositionally}, a finding with clear physical correlations.

It should be noted that with salience and attention-based approaches, there is a danger of over-interpretation, particularly in cases where physical explanations are searched for. There can be cases where salience maps produce the same explanation for \textit{all} classes, this can happen, for example, if the network is particularly responsive to edges in the input, as opposed to more meaningful features \cite{rudin2019stop}. There is a tendency in the literature to produce salience maps only for the top-ranked class, but to ensure that the network really is picking a class due to certain region it is important to consider what parts of the image activate for other classes that are not the top class. Another challenge of deep interpretation techniques is that they may lead to non explainable results. In these cases, a physical explanation of feature attribution may be difficult to infer as the model may be exploiting correlations in the data distribution that do not have a physical explanation or cause, a problem commonly known as shortcut learning \cite{robinson2021deep}. A combination of ML interpretations and secondary experimentation or simulation is advisable in these cases \cite{becareful2021}.

\section{Experimental predictions and explanations}

We build models in order to understand and guide experiments. Interpretability is a desirable characteristic of ML models in experimental contexts as it facilitates tasks such as characterization, optimization, sensitivity analysis and hypothesis testing. In materials science, physical models are often developed to approximate input-output relations in experimental processes and interpret, identify or optimize dominant physical parameters. Examples of such models include molecular dynamics simulations of carbon nanotube synthesis or drift-diffusion models of semiconductor devices. This category of models is derived from known mathematical relations with chosen inductive biases and is often designed with intrinsic experimental interpretability in mind. However, explanations of these models are subject to the tension between correctness, interpretability and completeness which we introduced in Figure ~\ref{fig:key-concepts}. 

We argue that a principled adaptation of ML models to the experimental context may preserve a useful degree of interpretability. For example, a combination of intrinsic physical parametrization and a surrogate ML model has been applied to the complex problem of mapping fabrication variables to final figures-of-merit in layered semiconductors (solar cells, transistors, etc.) \cite{ren2020embedding} or first-principles calculations have been used to constrain surrogate-based compositional optimization \cite{sun2021data}. In these hybrid models, ML enhances the model capacity of a physical parametrization to better approximate experimental data \cite{ren2020embedding}, account for uncertainty or deal with noise \cite{sun2021data}. Interestingly, it also allows the smooth integration of first-principles calculations into experimentation \cite{kusne2020fly,sun2021data}.

In scenarios where hypotheses are tested experimentally, predictive power may be second to induction or explainability. Traditionally, ML models with intrinsic interpretability have proven useful in this setting even if the absence of causal models, for example in the case of inference and explanation of degradation processes \cite{naik2021discovering}. Successful hypothesis testing in this context has traditionally relied on careful experimental design, expert heuristics and control of confounding factors. Novel ML techniques may relax these conditions and actively inform experimentation or inference \cite{pearl2018theoretical, kalinin2021describing}. ML models with a high degree of intrinsic interpretability have been used to identify dominant material descriptors in high dimensional material screening spaces \cite{vasudevan2019materials} or to actively guide experimental interventions with physical or causal constraints \cite{sun2021data, liu2021exploring}. In the same way, the capacity of ML models to approximate complex conditional distributions, may facilitate understanding of complicated physical and chemical systems for which high-performing physical simulations are limited, as demonstrated by recent advances in likelihood-free inference \cite{carleo2019machine}.

Extrinsic methods such as \SHAP and salience are proving powerful in coupling ML models to experimental procedures. \SHAP analysis has been used to understand ML models that predict the efficacy of organic capping layers for increasing stability of halide perovskites solar cells, highlighting the importance of low numbers of hydrogen bond donors and small topological polar surface ares \cite{hartono2020machine}. Salience methods have been used to identify the regions in 3D neutron spectroscopy signals that are most important for deciding the magnetic structure in a double perovskite, these regions are found to match with the regions identified by a trained physicist, but are found in a fraction of the time \cite{butler2020interpretable}. Salience methods were also used to identify the regions responsible for mis-classifications in an X-ray diffraction analysis deep neural networks, allowing human intervention where the model is likely to perform poorly \cite{oviedo2019fast}. 

These examples demonstrate how building interpretability is the key to successful application of ML for enhanced experimentation. Interpretable models not only increase the level of trust in the ML approach, but help to strengthen the relationship between the algorithms and the humans in the experimental loop. 

\section{PHYSICAL KNOWLEDGE BEYOND MODEL EXPLANATIONS}

Most applications of interpretability in the material science field have been driven by predictability goals, having interpretability as a second-order goal. We believe that some machine learning problems might be better framed with the explicit goal of extracting actual knowledge or causal interpretations \cite{kalinin2021describing, naik2021discovering, atkinson2019data, rudy2017data}. In exploratory scientific research, this often constitutes a better trade-off of the dimensions we explore in Figure~\ref{fig:key-concepts}. 

\begin{figure*}
   %\begin{center}
   \includegraphics[width=0.8\linewidth]{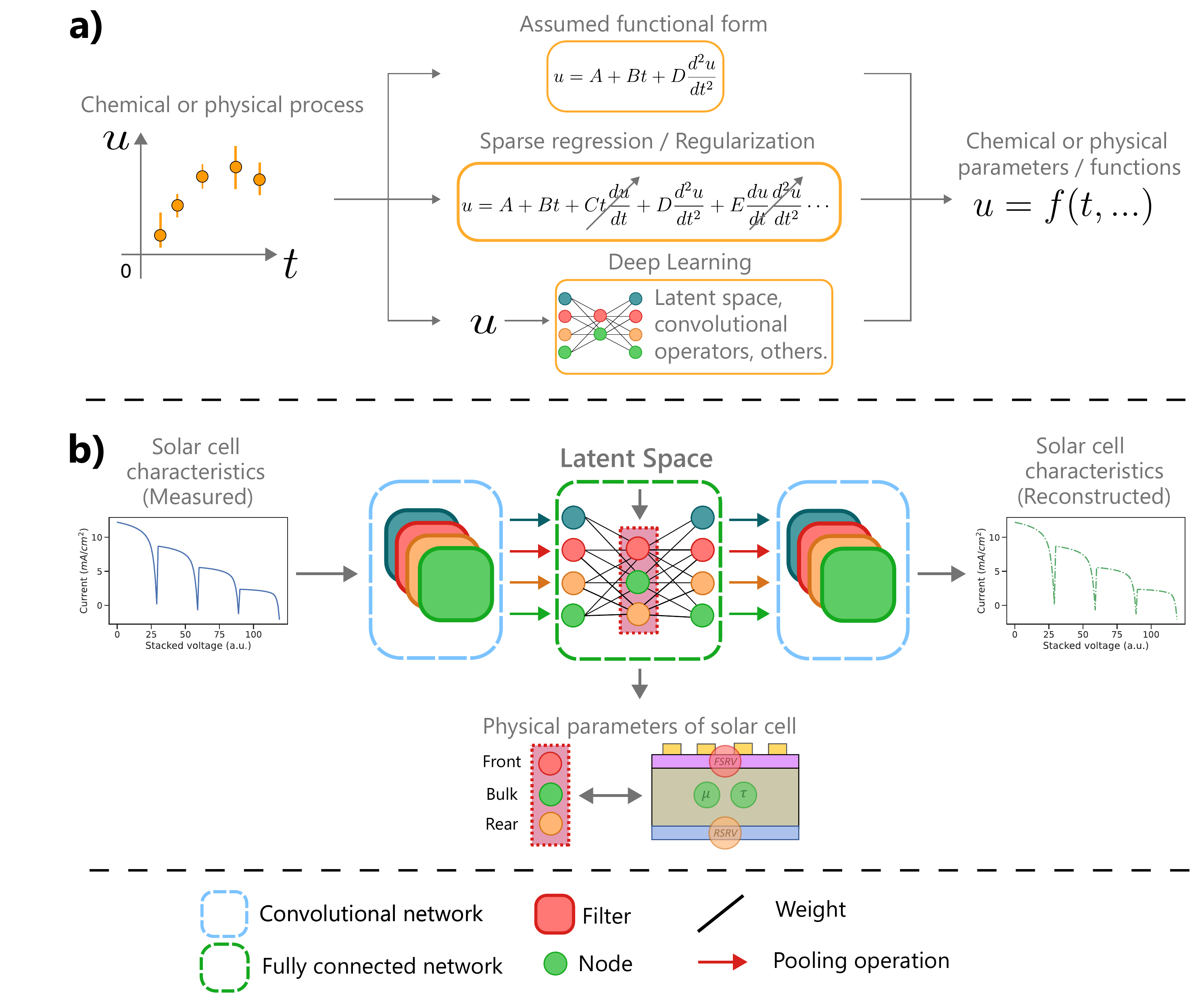}
   %\end{center}
   \caption{a) General taxonomy for the methods of direct knowledge extraction from physical or chemical data. b) Example of the application of deep auto-encoders to learn useful and interpretable representations of solar cell data. Current voltage characteristics of solar cells can be encoded in physical parameters, by analyzing these physical parameters we gain operational knowledge of solar cells and are able to infer path for device optimization \cite{ren2020embedding}. \label{fig:vae-nn}}
\end{figure*}

One approach in the broader physics community, summarized in Figure~\ref{fig:vae-nn}a, consists in designing or learning models that directly extract knowledge from noisy experimental data. An initial approximation to the problem is to assume an functional form and fit various coefficients to experimental data. A more robust approach consists in using sparse regression \cite{rudy2017data, naik2021discovering} or genetic algorithms \cite{atkinson2019data} to explore many potential functional forms and find those that better explain the data. An evolution of these techniques is based on deep learning methods, commonly on deep autoencoders (AEs), which for example have been shown successful in extracting order parameters for phase transitions \cite{wetzel2017unsupervised, walker2020deep}, explaining heat transfer phenomena \cite{he2020unsupervised} or disentangling physical phenomena in microscopy data \cite{ziatdinov2021robust, kalinin2021disentangling}. AEs work by learning to reconstruct an input, while passing through a reduced dimensional space, termed the latent space, thus learning compressed representations of the data, see Figure~\ref{fig:vae-nn}.  This latent space can be constrained by various methods, for example by adding penalty losses to the latent space that penalize for physical variables \cite{ren2020embedding}, explicitly defining hierarchical (or even causal) graphical structures in the latent space \cite{hsu2017unsupervised}, using adversarial training to constraint representations \cite{sarhan2019learning}, etc. Simple operations in this interpretable latent space, such as clustering or regression, make it possible to find physical insights or perform physically-relevant predictions. Figure~\ref{fig:vae-nn}b presents an example of using AEs to extract physically-relevant knowledge from noisy experimental data and a physical model, and use these learning to design an optimal solar cell fabrication process.

As another example, the so-called $\beta$-VAE\cite{higgins2016beta} introduces additional constraints to enforce orthogonality and sparsity on the latent space, so that the dimensions are uncorrelated and the VAE will only use the minimum number of dimensions required for reconstruction of the data. This kind of $\beta$-VAE type approach was recently shown to extract parameters that are interpretable as the driving parameters of ordinary differential equations from data of dynamic processes \cite{lu2020extracting}.

Another challenge in the field is related to the lack of confidence intervals or error distributions for model explanations. In contrast to classical statistical approaches on linear models, interpretations of modern machine models do not produce any notion of uncertainty. This fact greatly limits the confidence of any insights extracted from the model, as there is no inherent notion of uncertainty to them. Various works have explored uncertainty and bias in model explanations and have proposed ways to account for them \cite{li2020efficient, schwab2019cxplain}. We expect that future interpretability approaches in material science will integrate this inherent notions of uncertain into the insights extracted from ML models.

Finally, a continuing and fundamental challenge in ML interpretability is that explanations do not have strong causal guarantees or resilience against co-founding effects. Thus, the real-world insights gained from interpretability tend to be limited by the judgment of the scientist or secondary confirmation by experiments or simulation. The field of causal inference is witnessing a renaissance in fields of AI where explainable chains of action are legally necessary, such as autonomous vehicles. While most ML methods work on identifying correlations in data, they say nothing about cause and effect; the leading proponent of causal inference Judea Pearl has pronounced ML methods to be ``profoundly dumb'' for this reason~\cite{pearlbook, pearl2018theoretical}. This constitutes a very active area of research in mainstream machine learning, and we are optimistic about future progress in the field. By combining ML with the tools of causal inference it may be possible to learn new cause and effect relationships from materials data. A recent pioneering example suggesting the potential for this kind of approach was reported on electron microscopy data by Ziatdinov and colleagues\cite{ziatdinov2020causal} who report combining ML with causal inference to uncover mechanisms driving ferroelectric distortions, based on experimental micrographs. We believe that one great advantage of causality approaches in hard experimental science is that there is significant control of confounding factors by means of traditional experimental design.

These concepts of causal inference are also being explored in the context of generating extrinsic explanations of ML models. We have described how Shapley analysis provides a powerful, principled approach to asking what-if questions of models and producing insightful interpretations (see section~\ref{what_if} \textit{What-if interpretations}). However, these values can be susceptible to sampling unrealistic parameter combinations, for example the Shapley method may try to calculate the dielectric constant for a metal with a band gap of 2 eV, if metal/non-metal and band gap were separate input parameters. VAEs have been explored as a method for ensuring that sampled scenarios for arriving at final Shapley values fall within reasonable distributions~\cite{frye2020shapley}. Shapley values also have no concept of causality, so the density of a material may just as well result in composition as \textit{vice versa}. By relaxing the symmetry constrains for Shapley values it becomes possible to develop interpretations that respect known causal chains~\cite{frye2019asymmetric}. These developments have thus far only been applied in computer science, however their applicability of developing more robust and meaningful explanations for materials science ML is clear.

Other fields such as econometrics have a long tradition of developing statistical models to get insights about certain phenomena \cite{crown2019real, athey2015machine}. We imagine these approaches extracting meaningful, actionable information from complex materials data, such as processing conditions, complex compositional landscapes and large scale simulations. 

\section{Conclusions}

The technological advances of machine learning have been felt in all areas of science and technology in the past decade, from the original \imagenet moment \cite{russakovsky2015imagenet} to the recent successes of \alphafold \cite{jumper2021highly}. But as the initial excitement at these disruptive successes starts to fade, the long process of realising the true potential of ML for understanding the world begins. As with any largely empirically-developed technological advance, the process of understanding just why it works so well will only increase the breath and depth of the application of ML. In this paper, we have outlined some of our experience and observations of the nascent field of interpretable ML, in the context of materials science. The methods that we have outlined here cover interpretability for the range of ML methods that are becoming increasingly popular in materials science. Many of the methods we have presented are as easily implemented as the the ML models they interpret. As such we hope that in future every ML paper in materials science will include some efforts to understand the derived models and to extract more knowledge from the information. We have also tried to provide a balanced critique of the potential short-comings of these methods, interpretable ML is not a silver bullet for model understanding and constitute just an initial approximation to causal hypothesis generation. In the final analysis, to paraphrase David E. Womble (who may have been paraphrasing Max Planck), interpretable ML will not replace human experts, but human experts who embrace interpretable ML will replace those who don't.

\section{Acknowledgements}
We thank Professor Volker Deringer and Dr Noor Titan Putri Hartono for useful discussion. We thank Pedro Costa for his contributions to figure design. This worked was supported by the National Research Foundation (NRF), the Singapore Massachusetts Institute of Technology (MIT) Alliance for Research and Technology’s Low Energy Electronic Systems research program, Microsoft AI for Good.  

\section{Author Contributions}
KB and FO conceived the work and wrote the manuscript with key intellectuals contributions from TB and JLF.

\bibliographystyle{apsrev4-1}
\bibliography{bibliography} 

\end{document}